\documentclass[twocolumn,nofootinbib,prl,aps]{revtex4}
\usepackage{amsmath}
\usepackage{amsfonts}
\usepackage{amssymb}
\begin{document}
\title{Nonlinear Dynamics of Quasiclassical Spin Moment in a Swept Field}
\author{A.K. Zvezdin}
\thanks{This work was partially supported by RFBR (N99-02-17830) and INTAS
(N99-1839)} \affiliation{Institute of General Physics, Russian
Academy of Sciences, Vavilov st..38, 117942, Moscow, Russia}
\begin{abstract}
Quantum dynamics of anisotropic spin system with large spin moment
in a swept magnetic field is theoretically investigated. Magnetic
field of this type induces vortex static electrical field, that
breaks down the axial symmetry and induces new coherent quantum
phenomena: the occurrence of band energy spectrum with continuous
spin states, Bloch-type oscillations and interband Zener
tunnelling effect. These quantum phenomena display themselves, in
particular, in magnetization jumps and susceptibility peaks in the
investigated spin system.
\end{abstract}
\maketitle \sloppy {\bf 1.} The problems concerning to quantum
dynamics of spin systems have aroused considerable interest in
recent years (see references\cite{bar,fri,ses,dob,gar,har}). In
many respects it is interlinked with recent discoveries of the
macroscopic quantum tunnelling of magnetization
\cite{bar,fri,ses,dob,gar}, quantum hysteresis, molecular
bystability, novel type of magnetic oscillations associated with
Berri phase appearance. These mesoscopic phenomena were discovered
in so called ``giant spin moment systems'' -- in the magnetic
nanoclusters Mn$_{12}$, Fe$_{8}$ with large spin moment in the
ground state ($S=10$). Some mesoscopic spin phenomena are likely
to be discovered in the rare-earth ions with large angular
momentum (Dy$^{3+}$, Tb$^{3+}$, Ho$^{3+}$, Er$^{3+}$). Of
particular interest are problems pertaining to the macroscopic
quantum coherence, quantum measurements in spin systems and the
quantum correlations violation  due to interaction with
environment and especially quantum to classical behavior
transition. It should be stressed that the observing conditions of
these coherent phenomena are more strict than that for
above-listed. Therefore it is seemed to be of importance to
explore the novel situations when quantum coherence phenomena
appear. This problem is of practical concern for magnetic
nano-electronics (spintronics) and quantum computer science. One
of the most intriguing ideas in spintronics  is the
``giant-spin''-nanocluster usage as a bistable unit for further
generation of the molecular memory. These nanoclusters may be
interesting as a promising realization of qubits in quantum
computing \cite{los,dob2,zvezdin1,bur}.

The aim of the paper is quantum dynamics investigation of the
large spin moment quantum system in a swept magnetic field. Such a
field produces a torque on the spin system inducing the spin
precession and thus, displays new effects in the dynamics of the
spin system. The present paper develops the ideas surveyed by
author earlier as applied to the antiferromagnetic nanoclusters
\cite{zvezdin2}, metal rings and ring-like molecules
\cite{zvezdin3}.

{\bf 2.} We consider a quantum system (ion, molecule, cluster)
under the swept magnetic field. The Hamiltonian of the system is
taken to be
\begin{equation}\label{quant}
  \mathcal{H}=g\,\mu_{B}\,\vec{J}\,\vec{B}(t)+V_{CF},
\end{equation}
where $V_{CF}$ is the crystalline field operator. It is suggested
that  $J\gg 1$, thus in order to describe the quantum dynamics of
the investigated system the quasiclassical approach will be used.

The crystalline field is represented as
\begin{displaymath}
V_{CF}=V_{CF}^{0}+V'_{CF},
\end{displaymath}
where $V_{CF}^{0}$ is the crystalline field with ``easy plane''
symmetry and $V'_{CF}$ creates the anisotropy in plane; in so
doing $|V'_{CF}|\ll |V_{CF}^{0}|$. Let $z$ axis of the Cartesian
coordinate system be perpendicular to the ``easy plane'' which
contains 2-nd order axis selected as $x$-axis.\footnote{The
generalization for situations with the anisotropy in plane being
characterized by other symmetry elements, for example $4_{z}$,
$3_{z}$, $6_{z}$, is not a particular problem.}

Assume that the magnetic field is
$\vec{B}=\left(0,\,0,\,B_{z}\right)$ with $B_{z}$ being
time-dependent. Let $B(t)=B_{1}\,t/\tau$, where $B_{1}$ and $\tau$
are the characteristics of the field increase (decrease) process.
According to the well-known analogy let designate the
$j_{m}=B_{1}/4\pi\tau$ as a ``magnetic current''.

{\bf 3.} To describe the spin dynamics the coherent quantum states
technique $|\theta,\,\varphi\rangle$ \cite{per,fra} will be
employed with $\theta$ and $\varphi$ being polar and azimuthal
angles of the angular momentum.  These angles are measured from
the $z$ and $x$ axes respectively.

The Lagrangian of the system can be represented as:
\begin{equation}\label{lagr}
\mathcal{L}=-\frac{M}{\gamma}\,\left(cos\theta
-1\right)\,\dot{\varphi}-E,
\end{equation}
\begin{equation}\label{E}
\begin{split}
&E=-K_{1}\,\sin^{2}\theta
+K_{2}\,\sin^{2}\theta\,\sin^{2}\varphi-M B(t)\,\cos\theta,\\
&|K_{2}|\ll|K_{1}|,\,\,\,K_{1}>0,\,\,\,B||z.
\end{split}
\end{equation}
This Lagrangian can be deduced with conventional coherent quantum
states technique \cite{fra}. All terms in (\ref{lagr}) are of
apparent physical meaning. The first and the second terms in
(\ref{lagr}) are: the so called Wess-Zumino term (kinetic energy)
appeared because of the nonorthogonality of the coherent states at
$t$ and $t+\Delta t$ and the total energy $E$. The first and the
second terms in (\ref{E}) are the anisotropy energy (crystalline
field). The $\theta$ and $\varphi$ functional dependences of $E$
are chosen in the simplest but sufficient way to elucidate the
principle aspects of the problem. The last term in (\ref{E}) is
Zeeman energy. The Euler-Lagrange equations for Lagrangian
(\ref{lagr}) are equivalent to the Landau-Liphshits equations
(without attenuation).

{\bf 4.} The partition function of the quantum system can be
represented as the functional integral in the Eucledian space
($\tau=it$)
\begin{equation}\label{sum}
  \mathcal{Z}=\int D\cos\theta\,\int D\varphi\,e^{\frac{1}{\hbar}\int_{0}^{\hbar\beta}d\tau\,\mathcal{L}(\theta,\varphi)},
\end{equation}
where $\beta = \frac{1}{T}$, $\theta =\theta(\tau)$,
$\varphi=\varphi(\tau)$. As $|K_{2}|\ll |K_{1}|$ one can suppose
that in moderately high magnetic fields $\theta -\pi/2 \ll 1$.
Then integrating the partition function expression (\ref{sum})
upon $\theta$ yields \footnote{The calculation procedure can be
reduced to the following. It is not difficult to see that value
$\theta=\pi/2$ is the extremum point of the Eucledian action of
the functional
$S_{E}=\int_{0}^{\hbar\beta}d\tau\,\mathcal{L}(\theta ,\varphi)$
(at $B=0$). From the stationarity action condition (Saddle-point)
$\delta S_{E}=0$ it follows (at $B\neq 0$):
\begin{equation}\label{ax}
  \frac{\pi}{2}-\theta_{M}\approx\frac{M}{2K_{1}}\,\left(B(t)-\frac{i}{\gamma}\,\dot{\varphi}\right),
  \tag{$4'$}
\end{equation}
Substituting (\ref{ax}) to (\ref{sum}), then expanding $S_{E}$
about $\Delta\theta =(\theta-\theta_{M})$ to second order, and
calculating, if need be, the Gaussian integral upon
$\Delta\theta$, equations (\ref{sum2}) and (6) can be obtained. In
equation (\ref{sum2}) the Gaussian integral is omitted for
simplification of the formula because of being further
unessential.}:
\begin{equation}\label{sum2}
  \mathcal{Z}=\int\,D\varphi\,e^{\frac{1}{\hbar}\int_{0}^{\hbar\beta}d\tau\,\mathcal{L}_{eff}(\varphi)},
\end{equation}
where
\begin{equation}\label{L}
\mathcal{L}_{eff}=-\frac{I\dot{\varphi}^{2}}{2}-U_{A}(\varphi)-\gamma
I+iB(t)\dot{\varphi},
\end{equation}
where $I=\frac{M^{2}}{2K_{1}\gamma^{2}}$ is the moment of inertia
of the system, $U_{A}=K_{2}\,\sin^{2}\varphi$. The last term in
(\ref{L}) is defined as the Zeeman energy contribution.
\newline The Lagrangian (\ref{L}) can be presented (in the real time)
as following:
\begin{equation}\label{L1}
  \mathcal{L}_{eff}=\frac{I\left(\dot{\varphi}-\gamma
  B\right)^{2}}{2}-K_{2}\sin^{2}\varphi
  -\frac{\gamma^{2}IB^{2}}{2},
\end{equation}
which is convenient for the gauge transformation performance and
for making analogies with other quantum systems (the Josephson
junction \cite{ave,sch}, mesoscopic conducting ring or ring-like
molecule \cite{zvezdin3,eck}, antiferromagnetic nanocluster
\cite{zvezdin2}). One can readily see that the Lagrangian
(\ref{L1}) can be recast with an accuracy of the total time
derivative as (see also \cite{zvezdin4,zvezdin5})
\begin{equation}\label{L2}
  \mathcal{L}_{eff}=\frac{I\dot{\varphi}^{2}}{2}+\frac{K_{2}}{2}\,\cos2\varphi
  +\gamma I\dot{B}\varphi.
\end{equation}
It should be specially noted that the variable $\varphi$ is
defined here not on the $S^{1}$ $(0\leqslant\varphi <2\pi)$ set as
it usually assumed in the theory of angular momentum but on the
real numbers $\mathcal{P}$ set $(\varphi \in R^{1})$. The latter
in the present problem represents a product bundle of the $S^{1}$
set. $S^{1}$ plays a role of the base of this fiber set
$\mathcal{P}$. This remark is obviously important in the given
context as presence of the swept field $B_{z}(t)$ breaks down the
symmetry $\varphi\rightarrow\varphi+2\pi n$ transformation, where
$n$ is integer, therefore, $S^{1}$ set, commonly used in such
problems, must be enlarged up to the $\mathcal{P}$. The axial
symmetry violation is apparent because the swept field $B_{z}(t)$
generates the electrical axially symmetrical field
$\vec{E}_{\varphi}$,  by virtue of Maxwell equation
$-\frac{1}{c}\dot{\vec{B}}=curl\vec{E}$.

Systems with potential energy of the ``tilted washing board ''
type $U(x)=U_{0}(x)+cx$, where $U_{0}(x)$ is the periodical
function of $x$, $c$ is constant value, were already investigated
earlier. As for example it is possible to mention the electron
motion in electrical field in a crystal \cite{wan,bloch} or
Josephson junction dynamics when direct current flowing
\cite{ave,sch}. Therefore one can expect the manifestation of some
properties in the spin momentum dynamics which are similar to that
of above discussed systems. Such characteristic properties are the
band energy spectrum occurrence, the Bloch oscillations
\cite{wan,bloch} and the interband Zener tunnelling effect
\cite{zen,kel,eil}. Let's consider these issues in more detail.

{\bf 5.} The generalized momentum appropriate to the coordinate
$\varphi$ equals to $p_{\varphi}=\frac{\partial
\mathcal{L}_{eff}}{\partial\dot{\varphi}}=J\,\dot{\varphi}=I(\dot{\varphi}-\gamma
B)$. Then the Hamiltonian of the system
$\mathcal{H}=p_{\varphi}\dot{\varphi}-\mathcal{L}_{eff}$ can be
represented as
\begin{equation}\label{H}
  \mathcal{H}=\frac{1}{2I}\left(
  P_{\varphi}-\gamma I\,B\right)^{2}+U_{A}(\varphi),
  \end{equation}
where
$P_{\varphi}=\frac{\hbar}{i}\frac{\partial}{\partial\varphi}$.
\newline The gauge transformation
$\Psi_{1}(\varphi)\longrightarrow\Psi(\varphi)\,e^{\theta(\varphi,t)}$,
where $\theta=\frac{i}{\hbar}I\gamma B\,\varphi$ (the label ``1''
of the function $\Psi$ will be further omitted), transfers the
Schroedinger equation to the following type:
\begin{equation}\label{6}
i\hbar\dot{\Psi}=\left[\frac{\hat{p}_{\varphi}^{2}}{2I}+U_{a}(\varphi)-I\gamma\dot{B}\,\varphi\right]\,\Psi.
\end{equation}

At first let's consider the {\it eigenvalue problem} of the
Hamiltonian (\ref{H}) at $B=0$ . The eigenstates of the
Hamiltonian (\ref{H}) are the Bloch functions
\begin{equation}\label{bloch}
  \Psi_{s}(\varphi+2\pi)=e^{i\pi m}\Psi_{s}(\varphi),
\end{equation}
where $m$ is an arbitrary real number, $s$ is the energy band
number.  It would appear reasonable to designate the parameter $m$
as quasispin, (compare with the quasimomentum for a band
electron). By analogy with the term -- ``charge states''-- for
similar states in the theory of Josephson effect it is possible to
call the states (\ref{bloch}) as the ``continuous spin states''.
It is common knowledge for the spin moment component to be
quantized at the fixed direction. In the case being considered the
``quasispin'' is a arbitrary real number $(m \in
R^{1})$

The difference between these two types of states can be explained
as follows. The quantized spin states are specified at $S^{1}$ set
$(0\leqslant\varphi<2\pi)$, in so doing the spin moment
quantization is clearly associated with the axial symmetry of the
system, in other words, with the boundary conditions $\Psi
(\varphi +2\pi)=\pm\Psi(\varphi)$. The absence of this symmetry in
the dynamic symmetry group of the Lagrangian (\ref{lagr}) cancels
the quantization of a spin moment. Instead of this the
``continuous spin states'' i.e. Bloch functions (\ref{bloch}), are
achieved in the fiber set $\mathcal{P}$ $(\infty <\varphi
<\infty)$.

Let $U_{A}(\varphi)=-(1/2) \,K_{2}\,\cos 2\varphi$, where $K_{2}$
is a constant. Then the Schroedinger equation for the Hamiltonian
(\ref{H}) is reduced to the Mathieu equation from the theory of
which it follows that the energy spectrum of the Hamiltonian
(\ref{H}) has a band structure, i.e. the eigenvalues of (\ref{H})
$E_{n}(m)$ are the functions defined in the appropriate Brillouin
zones.  At $K_{2}\approx 0$ the band structure corresponds to the
approximation of ``free electrons'' type:
\begin{equation}\label{8}
  E_{s}(m)=\frac{\hbar^{2}m^{2}}{2I}
\end{equation}
with the forbidden bands at the boundaries of the Brillouin zones:
$m_{B}=s \,(s=\ldots-2,-1,0,1,2,\ldots)$ which are narrow in
accordance with $K_{2}\ll K_{1}$.

Near the Brillouin zone boundary, for example, near $m_{B}=-1$,
the wave function can be represented as follows
\begin{equation}\label{PSI2}
  \Psi(\varphi)=u(\varphi)\,e^{-iEt},
\end{equation}
where $u(\varphi)=A_{1}e^{im\varphi}+A_{2}e^{i(m+2)\varphi}$. The
Schroedinger equation for the Hamiltonian (\ref{H}) can be written
as (the Mathieu equation)
\begin{equation}\label{shr}
\begin{split}
  &u^{\prime\prime}+\left(\mu-2d\cos
  2\tilde{\varphi}\right)u=0,\\
\text{where}\;\;\; &\mu=\frac{2IE}{\hbar^{2}},\;
d=\frac{IK_{2}}{2\hbar^{2}}.
\end{split}
\end{equation}
Used here is a new variable $\tilde{\varphi} =\varphi + \pi/2$.
The sign ``$\sim$'' will be further omitted. Substituting
(\ref{PSI2}) to (\ref{shr}) yields
\begin{equation}\label{expr1}
  \begin{split}
  &\left(\mu-m^{2}\right)A_{1}-dA_{2}=0,\\
  &-dA_{1}+\left(\mu-(m+2)^{2}\right) A_{2}=0,
  \end{split}
\end{equation}
whence it follows that
\begin{equation}\label{expr2}
  \mu=\left(\frac{m^{2}+(m+2)^{2}}{2}\pm
  \sqrt{\frac{\big(m^{2}-(m+2)^{2}\big)^{2}}{4}+d^{2}}\right).
\end{equation}
In particular from the expression (\ref{expr2}) it is clear that
at $m_{B}=-1$ the forbidden band width (i.e. between allowed zones
with $s=0$ and $s=1$) equals
\begin{equation}\label{delta1}
  \Delta
  E_{10}=\,\;\;\frac{2\hbar^{2}d}{I}=K_{2}.
\end{equation}
The allowed zone widths of the zero, first and third zones are
$E_{0}\approx\frac{2\hbar^{2}}{I},\,E_{1}\approx\frac{6\hbar^{2}}{I}$
and $E_{2}\approx\frac{10\hbar^{2}}{I}$, respectively, whereas the
forbidden band widths rapidly decrease (at $K_{2}\ll
 K_{1}$) when going to the higher energy bands. Then
 \begin{equation}\label{w}
\begin{split}
  &\Delta
E_{21}\approx\frac{\hbar^{2}}{I}\,\left(\frac{IK_{2}}{2\hbar^{2}}\right)^{2},\\
&\Delta
E_{32}\approx\frac{\hbar^{2}}{16I}\,\left(\frac{IK_{2}}{2\hbar^{2}}\right)^{2},\\
&\Delta
E_{43}\approx\frac{\hbar^{2}}{576I}\,\left(\frac{IK_{2}}{2\hbar^{2}}\right)^{2}\,\,\,{\text
{etc.}}
\end{split}
\end{equation}
The equations (\ref{8})(\ref{shr}), (\ref{expr2}), (\ref{delta1})
determine with sufficient accuracy the energy spectrum in the
limit of first two Brillouin zones.

{\bf 6.} Let's discuss effects of the swept magnetic field
$B_{z}(t)$ effects that as mentioned above can be considered as a
classical gauge field in equations (\ref{H},\ref{6}). The last
term $\gamma I\dot{B}\varphi$ in equation (\ref{6}) is playing the
same role as the energy -- $eF\cdot x$ ($F$ characterizes the
electrical field and $x$ is the electron coordinate) in the
well-known problem of the dynamics of the Bloch electron in an
electrical field.

Let consider the momentum $p_{\varphi}$ dynamics in the case of
magnetic field varying adiabatically slowly:
\begin{equation}\label{11}
  \left | I\gamma\dot{B} \right|\ll K_{2}.
\end{equation}
To describe the dynamics of a spin under the action of ``magnetic
current'' $j_{m}=\dot{B}/4\pi$ let's consider a wave packet of the
Bloch functions (\ref{bloch}). Let $\bar{m}$ and $\bar{\varphi}$
be the mean values of the quasispin and the coordinate of the wave
packet center and the values  $\Delta m$, $\Delta \varphi$
$(\Delta m\cdot\Delta\varphi\sim1)$ determine the corresponding
uncertainties. Under influence of ``magnetic current'' $j_{m}$ the
wave packet formed at $t=0$ moves to the boundary (for example,
right, i.e.  $m_{B}=1$) of the Brillouin zone where Bragg
reflection takes place. Here the wave packet velocity changes its
sign to an opposite and the wave packet appears at the left
boundary ($m_{B}=-1$), after that it moves again to the right
boundary of Brillouin zone $(m_{B}=1)$ then reflects again and so
on. This process is called the Bloch oscillations. It is described
by the following equations for $\bar{m}$ and $\bar{\varphi}$ mean
values:
\begin{equation}\label{12}
  \begin{split}
  &\dot{\bar{m}}=\frac{I\gamma\,B_{1}}{\hbar\tau},\\
  &\dot{\bar{\varphi}}=\frac{1}{\hbar}\frac{\partial
  E_{s}(\bar{m})}{\partial \bar{m}}.
  \end{split}
\end{equation}
During this (adiabatic) process the system remains in the state
with the definite $s$ and its physical properties as, for example,
the magnetic moment, are the oscillating functions of time with
the frequency
\begin{equation}\label{13}
  f_{Bloch}=\frac{I\,\gamma\, B_{1}}{\hbar\,\tau}
\end{equation}
If the external magnetic field has the harmonic contribution i.e.
\begin{equation}\label{17}
  B=B_{1}t/\tau+w\,\sin2\pi\,ft,
\end{equation}
then the resonances on the frequences $f=f_{Bloch}$ and
$f=r\,f_{Bloch}$, where $r$ is a rational number, are available
(the Stark ladder-like resonances).

When the ``magnetic current'' increases
\begin{equation}\label{18}
\left|\gamma\,I\,\dot{B}\right|\gtrsim K,
\end{equation}
the Zener tunneling effect occurs between adjacent zones. In
particular, the probability of the tunnel transition in a unit
time between zones with $s=0$ and $s=1$ equals
\begin{equation}\label{19}
  g_{01}=f_{Bloch}\,e^{-\beta},
\end{equation}
where $\beta=\frac{\pi K_{2}^{2}\tau}{\hbar^{2}\gamma B_{1}}$,
$\gamma=\frac{e}{mc}$.

{\bf 7.} Let us consider the average magnetic moment behaviour of
the considered spin system qualitatively. The magnetic moment
$z$-axis component is
\begin{equation}\label{mz1}
  M_{z}=M\,\cos\theta\approx
  M\left(\frac{\pi}{2}-\theta\right)=\frac{M^{2}}{2K_{1}}\,\left(B_{z}-\frac{\dot{\varphi}}{\gamma}\right).
\end{equation}
Averaging (\ref{mz1}) with the corresponding wave function yields
\begin{equation}\label{mz2}
  \langle M_{z}\rangle
  =\chi_{\bot}\,\left(B_{z}-\frac{\langle\dot{\varphi}\rangle}\gamma\right),
\end{equation}
where  $\chi_{\bot}=\frac{M^{2}}{2K_{1}}$.

Two limiting cases have to be considered: a) $K_{2}=0$ therewith
$g_{n,n\pm 1}=1$ and b) $g_{01}=0$. The first case can be defined
as a free precession of the spin system and the second one is a
Bloch oscillations case.
\newline At $K_{2}=0$ expressions (\ref{12}) give
\begin{equation}\label{28}
  \langle\dot{\varphi}\rangle=\gamma\left(B(t)+c\right),
\end{equation}
where $c$ is a constant value determined by initial conditions.
Substituting (\ref{28}) to (\ref{mz2}) yields
\begin{equation}\label{29}
  \langle M\rangle =M_{0},
\end{equation}
where $M_{0}\equiv-\chi_{\bot}\,c$ is the system magnetic moment
at $t=0$. Thus, the distinctive property of the free precession
case is that the accelerated spin precession influenced by
increasing (decreasing) magnetic field screens the ion
paramagnetic susceptibility contribution ($\chi_{\bot}\,B_{z})$ so
the average magnetic moment is field undependent.

In the case of the Bloch oscillations the situation changes
drastically. Here the $\langle M_{z}\rangle$ versus $B_{z}$
dependence represents the sum of ``ordinary'' linear
$(\chi_{\bot}\,B_{z})$ term and the periodical curve
$\langle\dot{\varphi}\rangle$ with the period of
\begin{equation}\label{DB}
  \Delta
  B=B_{1}\left(\tau\,f_{Bloch}\right)^{-1}=\hbar/\gamma I.
\end{equation}
At $K_{2}\ll K_{1}$ this periodical function is rather close to
``Saw-type'' with the amplitude  $\Delta\dot{\varphi}\approx
2I/\hbar$:
\begin{equation}\label{p}
  \gamma^{-1}\langle\dot{\varphi}\rangle=\begin{cases}
  B,& 0\leqslant B\leqslant b,\\
  B-2b,& b\leqslant B<3b,\\
  B-4b,& 3b\leqslant B\leqslant 5b \\
  \text{etc.},
  \end{cases}
\end{equation}
where $b\equiv\frac{\hbar}{\gamma I}$.

Generally, the $M_{z}(B_{z})$ dependence contains the
characteristic features of both limiting processes.
\newline Actually, as it was shown above (see
(\ref{expr2})) at $K_{2}/K_{1}\ll 1$ the forbidden band $\Delta
E_{21}$ width falls far short of  $\Delta E_{10}$, $\Delta
E_{32}\ll\Delta E_{21}$ etc. Therefore as a first approximation it
is possible to put $g_{12}=g_{23}=g_{34}=\ldots =1$, i.e. to
neglect the Bloch oscillations in the first and following excited
bands. It means that the precession in all bands except ground one
(i.e. $s=0$) can be considered as free. The  precession influenced
by $B_{z}(t)$ as a whole can be assumed as following: the wave
packet formed at $t=0$ slightly spreading out reaches the edge of
a Brillouin zone, then partially reflects with probability
$1-g_{01}$ and partially tunnels (with probability $g_{01}$) to
the next zone where it freely precessing. With this considerations
and the equations (\ref{19}),(\ref{p}) the $M_{z}(B)$ dependence
can be presented as follows:
\begin{equation}\label{MM}
  \chi_{\bot}^{-1}M_{z}=\begin{cases}
  0,& 0\leqslant B<b,\\
  2b(1-p),& b\leqslant B\leqslant 3b,\\
  4b(1-p)^{2},& 3b\leqslant B\leqslant 5b, \\
  6b(1-p)^{3},& 5b\leqslant B\leqslant 7b \\
  \text{etc.},
  \end{cases}
\end{equation}
where $p=e^{\beta}$, $\beta$ is determined in (\ref{19}). It is
easy to verify that the magnetic moment step $\Delta M_{z}$
magnitudes decrease as the number of the step increases. So, at
$B=b$ $(\Delta M_{z})_{10}=\chi_{\bot}\,2b(1-p)$; at $B=3b$
$(\Delta M_{z})_{21}=\chi_{\bot}\,2b(1-p)(1-2p)$; at $B=5b$
$(\Delta M_{z})_{32}=\chi_{\bot}\,2b(1-p)^{2}(1-3p)$ etc. This
means that the $M_{z}(B)$ dependence tends to saturation when $B$
increases.

All above mentioned considerations are concerning to the case of
$T=0$ K. Obviously, the thermal fluctuations (at $T\neq 0 $K) and
dissipation (i.e. dissipative environment interaction) breaks down
the studied quantum coherent effects. Taking into account the case
of finite temperatures is worth another look. Here we shall
restrict by only pointing out that the ``$T=0$ K''--theory is
valid when $T\ll K_{1}$ and $T_{2}\gg b\tau/B_{1}$, where $T_{2}$
is spin relaxation time of the system. At $K_{1}\approx$1 cm$^{-1}
$\footnote{It is worth to note that equation (\ref{E}) looks
realistic for a triplet ground state originating from the
crystalline field splitting of the ground multiple of the spin
system.} and $B_{1}/\tau\approx 10^{4}$ Oe/s we have $T\ll$2 K,
$T_{2}\gg$0.04 s. These limitations appears as quite executable
for modern low temperature experiments.

It should be stressed that the neglected above Bloch oscillations
in the bands $s=1$, $s=2$, etc. and the Bragg reflections at the
corresponding Brillouin boundaries give the additional periodical
terms into the $M_{z}(B)$ and $\chi(B)$ dependences. So, the Bragg
reflections at the boundary ($s=1,\,\,s=2$) give the additional
intermediate peaks of $\chi$ at the fields $B=2b,\,4b$ etc. The
reflection at the boundary ($s=-1,\,\,\,s=0$) decreases the peak
amplitude at the field $B=3b$. The Zener tunnelling at the
boundary ($s=-1,\,\,\,s=0$) even more complicates this picture due
to an interference  of the tunnelling wave function
($s=-1\longrightarrow s=0$) with the main wave function ($s=0$).
These questions are worth to be further investigated in more
detail.

In conclusion, it was shown that the external time dependent
magnetic field induces new coherent quantum effects of the
anisotropic spin system dynamics. They are: the occurrence of band
energy spectrum with continuous spin states, quasibloch
oscillations and interband Zener tunnelling effect. These quantum
effects manifest themselves as magnetizaion jumps and
susceptibility peaks in the investigated spin system.


\end{document}